\documentclass[aps,prl,twocolumn,nobibnotes]{revtex4-2}
 
\usepackage{amsmath}
\usepackage{amssymb}
\usepackage{graphicx}
\usepackage{xcolor}
\usepackage[colorlinks=true,citecolor=blue,linkcolor=blue,urlcolor=blue]{hyperref}
\usepackage{comment}
\usepackage{docmute}
\makeatletter
\def\@hangfrom@section#1#2#3{\@hangfrom{#1#2}#3}
\renewcommand{\@seccntformat}[1]{\csname the#1\endcsname.\ }
\makeatother

\usepackage{xspace}
\newcommand{\K}{$^{39}$K\xspace}
\newcommand{\Rb}{$^{87}$Rb\xspace}
\newcommand{\He}{$^{3}$He\xspace}
\newcommand{\Ne}{$^{21}$Ne\xspace}
 
\begin{document}
 
\title{Correlated comagnetometry for precision measurements}
 
\author{Yossi Rosenzweig}
\email[Contact author: ]{yosefros@post.bgu.ac.il}
\affiliation{Department of Physics, Ben-Gurion University of the Negev, Beer-Sheva, Israel}

\author{Yevgeny Kats}
\email[Contact author: ]{katsye@bgu.ac.il}
\affiliation{Department of Physics, Ben-Gurion University of the Negev, Beer-Sheva, Israel}

\author{Eli Sarid}
\affiliation{Department of Physics, Ben-Gurion University of the Negev, Beer-Sheva, Israel}

\author{Menachem Givon}
\affiliation{Department of Physics, Ben-Gurion University of the Negev, Beer-Sheva, Israel}

\author{Yonathan Japha}
\affiliation{Department of Physics, Ben-Gurion University of the Negev, Beer-Sheva, Israel}
 
\author{Ron Folman}
\affiliation{Department of Physics, Ben-Gurion University of the Negev, Beer-Sheva, Israel}
  
% -----------------------------------------------------------------------
\begin{abstract}
Magnetometers are among the most widely used probes in science and technology. Comagnetometers increase sensitivity by self-cancellation of magnetic noise, but only at low frequencies. We suggest a correlated measurement of two alkali species in one cell to cancel the magnetic background also at high frequencies. The inter-species phase difference of the light-matter interaction response function is found to be calibration free and insensitive to common-mode intensity noise. Utilizing a dark-matter signal as a testcase, the method achieves a thirtyfold background suppression, raising the signal-to-noise ratio by an order of magnitude or more, depending on the coupling to the different subatomic particles. We show that the method also provides model differentiation. The higher sensitivity and model differentiation open a path to novel probes for precision measurements in general and exotic fields in particular.
\end{abstract}
 
\maketitle
 
% -----------------------------------------------------------------------
%\section{Introduction}
% -----------------------------------------------------------------------
 
Optically pumped magnetometers can reach sub-femtotesla sensitivity~\cite{Kominis2003,Budker2007} and serve applications from biomagnetism to fundamental physics. In practice, however, the signal-to-noise ratio (SNR) of a shielded magnetometer is rarely set by its intrinsic noise: it is set by the residual magnetic background (shield Johnson noise, coil and current noise), which exceeds the intrinsic floor by more than an order of magnitude~\cite{Kornack2007,LeeRomalis2008}. An innermost ferrite shield partially mitigates the Johnson noise passively~\cite{Kornack2007}, but leaves coil, current, and other external noise in place. Comagnetometers address this by self-compensation: the polarized noble gas in an alkali--noble-gas cell cancels slow magnetic perturbations~\cite{Allred2002,Kornack2002}. This cancellation, however, is limited to low frequencies, below the noble gas Larmor frequency; above it the device is an ordinary, background-limited magnetometer. In this Letter we introduce a correlated measurement of two alkali species held in a single cell, and theoretically demonstrate that it cancels the correlated (i.e., common-mode) magnetic background at high frequencies as well, recovering the uncorrelated readout-noise floor. 

As we will discuss in the outlook, the method lends itself to a variety of applications. In our testcase example below, the correlated background amplitude is suppressed by a factor of $\sim 30$, allowing the SNR for a non-magnetic spin interaction to improve by up to an order of magnitude or more, depending on the signal’s coupling to the two species. In addition, we simulate a dark-matter transient, and show that the correlated signal also provides a calibration free readout of the signal's coupling structure that no single channel possesses.

Our testcase is built around a dark-matter model involving ultralight pseudoscalar (axionlike) particles (ALPs)~\cite{Kimball:2023vxk,Adams:2022pbo,ParticleDataGroup:2026aaa}. Because of the high occupation numbers required to account for the dark matter abundance, they behave as an oscillating classical field. Such fields can generically couple to the axial-vector currents of Standard Model fermions~\cite{Graham:2013gfa}. In the nonrelativistic limit this acts on nuclear and electronic spins like a magnetic field, but with couplings not proportional to the magnetic moments. Alkali--noble-gas magnetometers and comagnetometers are therefore leading dark-matter detectors~\cite{Pustelny:2013rza,Afach:2018eze,Bloch:2019lcy,Padniuk:2021dtr,Afach2023}, used to search for oscillations of the field associated with a galactic dark matter halo~\cite{Bloch:2019lcy,Lee:2022vvb,Bloch:2022kjm,Wei:2023rzs,Xu:2023vfn,Gavilan-Martin:2024nlo}, a solar dark matter halo~\cite{Wilson:2025lhq}, and transients~\cite{Afach:2021pfd,Khamis:2024oqa}. If such a signal is detected, disentangling its couplings to protons and neutrons has so far required comparing several distinct devices~\cite{Rosenzweig2024}. The correlated single-cell measurement introduced here addresses both needs at once: it removes the magnetic background that limits every single channel, and it reads out the proton--neutron coupling ratio from one vapor cell.

% ----------------------
%\section{The Dual-Alkali Correlated Measurement}

\emph{The Dual-Alkali Correlated Measurement.} The central observation of this Letter is that two same-cell alkali species in a single comagnetometer provide two independent readouts of the same field environment. \Rb, \K, and \He are polarized together via hybrid spin-exchange optical pumping~\cite{Babcock2003,Babcock2005,Walker1997}. The alkalis share the magnetic field and the electronic gyromagnetic ratio $\gamma_e$, including the nuclear slowing-down factor. A magnetic field therefore acts on both alkalis via an almost identical coupling. An exotic field, by contrast, couples to each species through its own combination of nucleon couplings, set by the fractional contributions of proton and neutron spins to the total nuclear spin~\cite{Kimball2015}. Because the spin content differs across \Rb, \K, and \He, one exotic field acts on the three species through three distinct effective couplings, parameterized by the ratio $\mathcal{R} \equiv \xi_n / \xi_p$ of the neutron to proton couplings ($\mathcal{R} = 1$: equal couplings; $\mathcal{R} = 0$: pure proton; any real value is possible in principle).

The common magnetic transfer function makes a differential readout an excellent common-mode rejection filter for the correlated magnetic background, in principle at all frequencies within the linear regime. The cancellation does not rely on the noble-gas self-compensation mechanism~\cite{Allred2002,Kornack2002}; it relies only on the two same-cell alkalis seeing one field. It therefore extends the rejection of magnetic noise to frequencies above the compensation point, where the single-alkali comagnetometer no longer attenuates. Consequently the amplitude difference of the two channels, each converted to magnetic-field units, cancels the correlated background itself, converting the background-limited SNR into a readout-noise-limited one, thereby enhancing the sensitivity. A second, complementary observable, is the inter-species phase difference $\Delta\varphi$ which is calibration free and insensitive to common-mode gain and intensity fluctuations, and carries the coupling ratio $\mathcal{R}$. We quantify these observations below.

%\section{Experimental realization}

\begin{figure}
    \centering
    \includegraphics[width=1\linewidth]{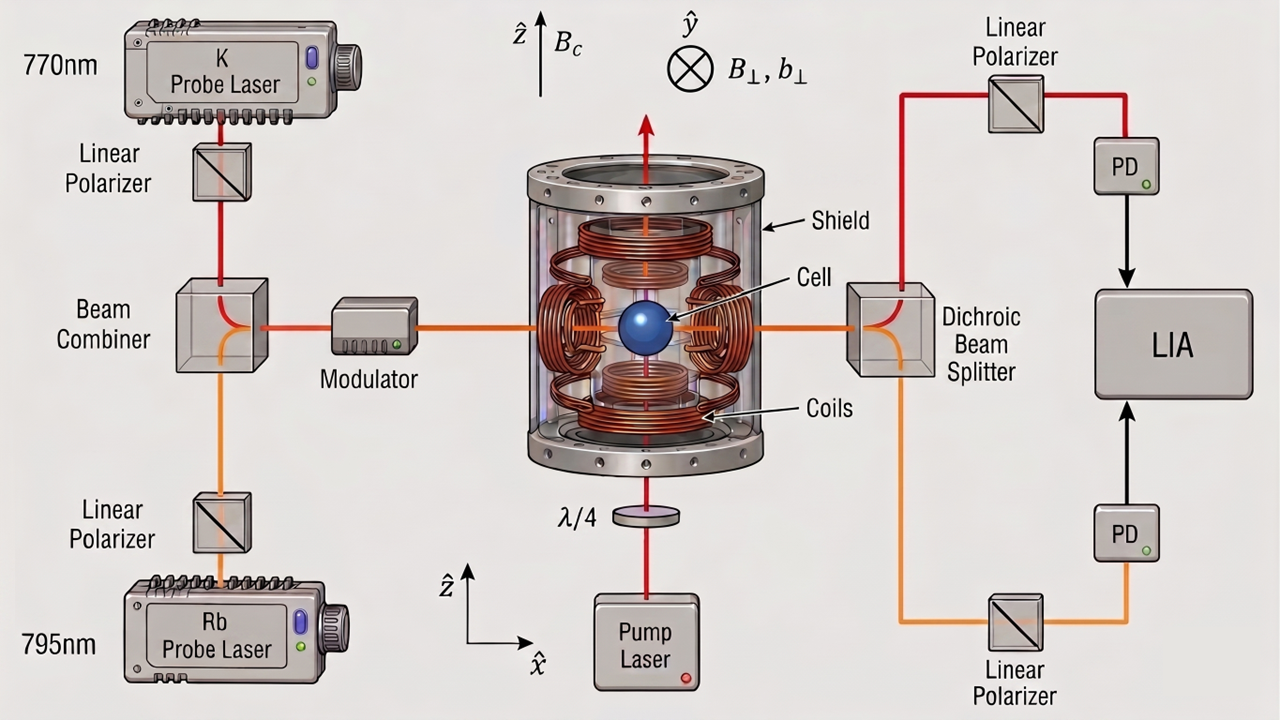}
    \caption{Experimental setup. A circularly polarized pump polarizes \Rb\ along $\hat{z}$ and, via spin exchange, \K\ and \He. Two overlapping probes near the \Rb\ and \K\ D1 lines traversing the cell along $\hat{x}$ are separated by a dichroic beamsplitter, and each is read out through a polarizer and photodiode (PD), demodulated by a shared lock-in amplifier (LIA) at the modulation frequency. The longitudinal field $B_z$ is tuned by the coils to the compensation point [Eq.~\eqref{eq:bz_alpha}]. Both the magnetic and exotic transverse perturbations, $B_\perp$ and $b_\perp$, respectively, point along $\hat{y}$. Auxiliary components are omitted for clarity.}
\label{fig:setup}
\end{figure}

\emph{Experimental Realization.} Both observables can be implemented in a standard comagnetometer apparatus (Fig.~\ref{fig:setup}). We take both the magnetic and exotic perturbation fields, $B_\perp$ and $b_\perp$, respectively, to lie along $\hat{y}$. A vapor cell filled with \He, \K, isotopically enriched \Rb, and N$_2$ (for quenching) is placed inside a mu-metal shield with magnetic coils, and heated to its operating temperature~\cite{SM}. A circularly polarized pump laser tuned to the \Rb\ D1 resonance polarizes \Rb\ directly and \K\ and \He\ by spin exchange~\cite{Walker1997, Klinger2023} along $\hat{z}$. The two alkalis are probed simultaneously by two overlapping, linearly polarized beams, each detuned ${\sim}0.1$\,THz to the blue of its own D1 line, traversing the cell along $\hat{x}$. A modulation is added to the probe signal to avoid measuring near the high amplitude region of the $1/f$ noise; the modulator could be a photoelastic modulator~\cite{duan2015light} or a high-frequency modulation of the shield coils~\cite{Gavilan-Martin:2024nlo}, with $\omega_{\mathrm{mod}}/2\pi$ in the tens of kHz, far above the detection bandwidth. After the probes traverse the cell and acquire Faraday rotation, they are split by a narrow-band dichroic beamsplitter followed by an interference filter in each arm, which makes the optical cross-talk between the channels negligible~\cite{SM}, and each polarization is measured by a polarimeter, which is then demodulated by a single multi-channel lock-in amplifier referenced to the shared modulation frequency $\omega_{\mathrm{mod}}$, so the two channels share a phase origin and any correlated electronic drift cancels.

For each species $j \in \{$\Rb, \K$\!\!\}$, the in-phase lock-in output returns the Faraday-rotation angle~\cite{Seltzer2008},
\begin{equation}
\theta_F^j(t) \;\propto\; \ell\, n_j\, r_e c\, f_j\, \mathcal{D}_j(\nu_j)\, P_x^j(t),
\label{eq:lockin}
\end{equation}
where $\ell$ is the cell length, $n_j$ the alkali density, $r_e$ the classical electron radius, $c$ the speed of light, $f_j$ the D1 oscillator strength, $\mathcal{D}_j(\nu_j)$ the dispersive lineshape factor at the probe detuning $\nu_j$, and $P_x^j(t)$ the transverse polarization projected onto the probe axis. In a search at an a priori unknown frequency $\omega \ll \omega_{\mathrm{mod}}$, each $\theta_F^j(t)$ is acquired as a time series and Fourier transformed,
\begin{equation}
\tilde{\theta}_F^j(\omega) \;=\; \mathcal{F}\{\theta_F^j(t)\} \;\propto\; P_x^j(\omega;\mathcal{R}),
\label{eq:theta_Px}
\end{equation}
preserving magnitude and phase. Since the proportionality factor in Eq.~\eqref{eq:lockin} is real and positive when the two probes are detuned to the same side of their respective D1 lines (opposite-side detuning adds a known constant $\pi$ to $\Delta\varphi$), $\varphi_j(\omega; \mathcal{R}) \equiv \arg \tilde{\theta}_F^j = \arg P_x^j(\omega;\mathcal{R})$, and the phase observable is
\begin{equation}
\Delta\varphi(\omega; \mathcal{R}) \;\equiv\; \varphi_{\mathrm{K}}(\omega; \mathcal{R}) - \varphi_{\mathrm{Rb}}(\omega; \mathcal{R}).
\label{eq:dphi_obs}
\end{equation}
Multiplicative factors in the detected signal that are static over an analysis record (probe intensity, photodetector responsivity, and the atomic prefactor of Eq.~\eqref{eq:lockin}) drop out of each phase $\varphi_j$, so $\Delta\varphi$ is calibration free.

% -----------------------------------------------------------------------
%\section{Coupled Bloch dynamics}

\emph{Coupled Bloch Dynamics.} The polarizations $\mathbf{P}_j$ ($j = $\;\Rb, \K, \He) evolve under the coupled Bloch equations~\cite{Padniuk:2021dtr,Rosenzweig2024,Padniuk:2023uyv},
\begin{align}
\dot{\mathbf{P}}_j &= \frac{1}{q_j}\!\bigg\{\!\bigg[\gamma_j\!\Big(\mathbf{B} + \!\sum_{k \neq j} \lambda_{jk} M_0^k \mathbf{P}_k\Big) + \xi_j \mathbf{b}\bigg] \!\times \mathbf{P}_j \nonumber\\
&\quad + \sum_{k \neq j} \kappa_{jk} n_k (\mathbf{P}_k - \mathbf{P}_j) - R_{pj}(\mathbf{P}_j - \hat{\mathbf{s}}) - \Gamma_j \mathbf{P}_j\bigg\},
\label{eq:bloch}
\end{align}
where $\gamma_j$ is the gyromagnetic ratio (electronic for the alkalis, nuclear for \He), $q_j$ the nuclear slowing-down factor (unity for \He), $\mathbf{B}$ the applied magnetic field, $\mathbf{b}$ a transverse exotic pseudomagnetic field with species-specific coupling $\xi_j$, $\lambda_{jk} M_0^k \mathbf{P}_k$ the Fermi-contact field induced on species $j$ by the magnetization of species $k$~\cite{Schaefer1989}, $\kappa_{jk}$ the spin-exchange rate coefficient, $n_k$ the density of $k$, $R_{pj}$ the pump rate along $\hat{\mathbf{s}}$, and $\Gamma_j$ the remaining relaxation. Only \Rb\ is pumped; \K\ polarizes through the fast \Rb--\K spin exchange and both alkalis polarize the \He, much more slowly~\cite{Babcock2003}. The exotic couplings are expressed through the nucleon couplings and the nuclear spin content $\sigma_{n,p}^j$~\cite{Kimball2015,Friar:1990vx,Engel:1989ix,Engel:1995gw,Flambaum:2006ip} as
\begin{equation}
\xi_j = \eta_j(\sigma_n^j \xi_n + \sigma_p^j \xi_p)\,,
\label{eta}
\end{equation}
where $\eta_j = q_j - 1$ for the alkalis and $1$ for the noble gas. Equation~\eqref{eta} assumes a negligible electron-spin coupling; exotic fields coupling to electron spins are also partially screened by magnetic shields, while nucleon couplings are not~\cite{jackson2016magnetic}. Parameter values are provided in Ref.~\cite{SM}.

Equation~\eqref{eq:bloch} admits a steady longitudinal solution along $\hat{z}$, set by the pump and the spin-exchange network. In a single-alkali comagnetometer the longitudinal field is tuned to the self-compensation point~\cite{Allred2002,Kornack2002}; with two polarized alkalis this cannot be met for both, since $\lambda_{\mathrm{Rb,He}} \neq \lambda_{\mathrm{K,He}}$, so we parameterize the compensation field as a weighted compromise,
\begin{align}
B_z(\alpha) = {} & -\bigl[(1-\alpha)\,\lambda_{\mathrm{Rb,He}} + \alpha\,\lambda_{\mathrm{K,He}}\bigr]\, M_0^{\mathrm{He}} P_z^{\mathrm{He}} \nonumber\\
         & -\lambda_{\mathrm{Rb,He}}\, M_0^{\mathrm{Rb}} P_z^{\mathrm{Rb}}
           -\lambda_{\mathrm{K,He}}\, M_0^{\mathrm{K}} P_z^{\mathrm{K}},
\label{eq:bz_alpha}
\end{align}
with $\alpha = 0$ ($\alpha=1$) recovering \Rb-optimal (\K-optimal) compensation. The fast \Rb--\K\ spin exchange locks the alkalis into a single collective transverse mode, fixing the optimal $\alpha$ near the \K\ fraction of the alkali density. We adopt $\alpha = 0.578$ for our cell~\cite{SM}. Together with the cell parameters it also sets the antiresonance location below. The transverse linear response at frequency $\omega$ is evaluated numerically following the fast scheme of Ref.~\cite{Rosenzweig2024}.

\begin{figure}[t]
\centering
\includegraphics[width=\columnwidth]{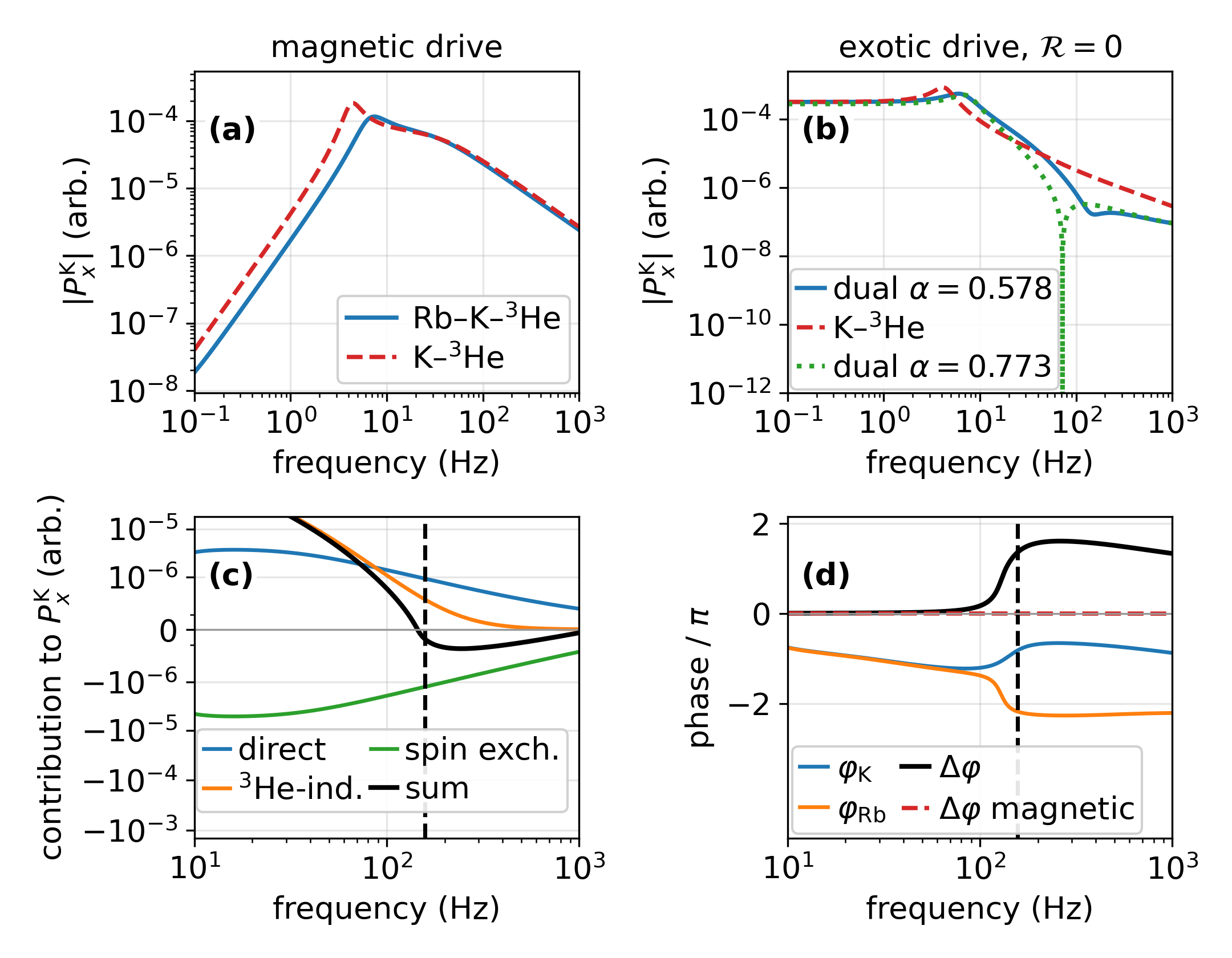}
\caption{Physics of the correlated dual-alkali response. (a),(b)~Absolute steady-state transverse response $|P_x^{\mathrm{K}}|$ to a perturbation along $\hat{y}$. (a)~Magnetic drive: the dual-alkali \Rb--\K--\He\ cell (solid blue) and a reference \K--\He\ single-alkali cell (red dashed) both show the standard low-frequency self-compensation. (b)~Exotic drive with $\mathcal{R}=0$: self-compensation no longer holds at low frequency, and the dual-alkali cell develops a sharp antiresonance near $157$\,Hz at $\alpha=0.578$ (solid blue). Changing only $\alpha$ to $0.773$ (green dotted; a $B_z$ change of 2.6\,nT, i.e., $1.3\%$) moves it to $\approx72$\,Hz, onto the perfect-cancellation locus where $|P_x^{\mathrm{K}}|$ nulls completely~\cite{SM}. The \K--\He\ single-alkali reference (red dashed) has no antiresonance at this coupling ratio~\cite{SM}. (c)~Mechanism at $\alpha=0.578$, $\mathcal{R}=0$: the three contributions to the \K\ response: the direct exotic drive (blue), the \He-induced field (orange), and the alkali--alkali spin exchange (green), shown as complex phasors projected onto the direct-drive phasor. Near the dip (dashed line) the opposing spin-exchange term cancels the direct-plus-\He\ drive and the in-phase sum (black) crosses zero; the quadrature component keeps the dip a finite minimum. (d)~Phases $\varphi_{\mathrm{K}}$ (blue), $\varphi_{\mathrm{Rb}}$ (orange), and $\Delta\varphi$ (black) for the exotic drive: each alkali's phase winds rapidly through its own antiresonance, in opposite senses, so $\Delta\varphi$ accumulates a single sharp swing. Crucially, neither feature appears in the magnetic drive: panel (a) shows no antiresonance in the response, and panel (d) shows a small $\Delta\varphi$ (red dashed) at the milliradian level across the band~\cite{SM}.}
\label{fig:fig2}
\end{figure}

Figure~\ref{fig:fig2} presents the physics of the correlated measurement. Panels (a) and (b) contrast the \K\ response of the dual-alkali cell with a reference \K--\He\ single-alkali cell. Both attenuate magnetic drives at low frequency, but they differ sharply for an exotic drive: for $\mathcal{R}=0$ the dual cell develops an antiresonance in the response $|P_x^{\mathrm{K}}|$ near $157$\,Hz that the \K--\He\ cell does not have at this coupling ratio. Quite generally, with $\xi_j=\eta_j(\sigma_n^j\xi_n+\sigma_p^j\xi_p)$ and $\eta_j>0$, an indirect drive opposes the direct one only when the two species' couplings differ in sign, i.e., when $\mathcal{R}$ lies between their sign-reversal ratios $-\sigma_p^j/\sigma_n^j$. At $\mathcal{R}=0$ this excludes \K--\He\ (\K\ and \He\ share coupling sign) but not \Rb--\K--\He, where \Rb\ supplies the opposing arm. The values of $\mathcal{R}$ for which each alkali--noble pair supports an antiresonance are tabulated in Ref.~\cite{SM}. Panel (c) resolves the \K\ response into its three exotic channels: the direct drive, the \He\ back-action (whose Fermi-contact return fails to cancel because $\lambda_{\mathrm{Rb,He}} \neq \lambda_{\mathrm{K,He}}$), and the alkali--alkali spin exchange. Their in-phase cancellation produces the antiresonance dip. The dip becomes an exact null on a one-dimensional locus in $(\alpha, \mathcal{R}, \omega)$ where the two real conditions $|P_+| = |P_-|$ and $\arg[P_+/(P_-)^*]=\pi$ hold simultaneously~\cite{SM}; the $\alpha=0.773$ curve in panel (b) sits on this locus at $72$\,Hz. Panel (d) shows the resulting phase anatomy: each alkali's phase winds through its own antiresonance in opposite senses, so the difference $\Delta\varphi$ accumulates a single sharp swing, while for a magnetic drive $\Delta\varphi$ stays at the milliradian level across the band. The location of the structure depends on $\mathcal{R}$ but also strongly on $\alpha$, providing an experimental handle through $B_z$.

% -----------------------------------------------------------------------
%\section{Dual-channel noise cancellation}

\emph{Dual-Channel Noise Cancellation.} We now use the dual channels to cancel the correlated magnetic background and quantify the SNR gain. The starting point is the Fourier-transformed lock-in signal of each channel, $\tilde{\theta}_F^j(\omega)$ [Eq.~\eqref{eq:theta_Px}]. We convert each channel to magnetic-field units by dividing out its magnetic transfer function $T_j^B(\omega)$, the complex response of $\tilde{\theta}_F^j$ to a known applied field, measured once by driving a calibrated field with the system's coils and forming $T_j^B = \tilde{\theta}_F^j/\tilde{B}$ at each frequency:
\begin{equation}
x_j(\omega) \equiv \frac{\tilde{\theta}_F^j(\omega)}{T_j^B(\omega)} = B(\omega) + \chi_j(\omega)\, b(\omega) + \frac{N_j(\omega)}{T_j^B(\omega)},
\label{eq:referred}
\end{equation}
with $\chi_j \equiv T_j^b/T_j^B$ the exotic response in field units and $N_j$ the uncorrelated readout noise. The common optical prefactor of Eq.~\eqref{eq:lockin} multiplies $\tilde{\theta}_F^j$ and $T_j^B$ alike and cancels in the ratio, so no absolute optical calibration is needed and the true field $B$ enters both channels with coefficient one. The correlated background is removed by an adaptive subtraction $r = x_{\mathrm{K}} - \hat{H}\,x_{\mathrm{Rb}}$, where $\hat{H}(\omega) = \langle x_{\mathrm{K}} x_{\mathrm{Rb}}^*\rangle/\langle |x_{\mathrm{Rb}}|^2\rangle$ is the least-squares (Wiener) transfer from the \Rb\ to the \K\ channel, a spectral ratio read off exotic-signal-free records rather than a fitted model~\cite{SM}. Because $\hat{H}$ is measured, the cancellation is self-calibrating: any error in the transfer functions $T_j^B$ folds into $\hat{H}$ and cancels, entering only the amplitude scale of the recovered signal. The residual $r = (\chi_{\mathrm{K}} - \chi_{\mathrm{Rb}})\, b$ plus uncorrelated noise retains the exotic signal with $g(\omega) \equiv |\chi_{\mathrm{K}} - \chi_{\mathrm{Rb}}|/|\chi_{\mathrm{K}}|$ relative to the single \K\ channel, while the correlated background cancels to the uncorrelated readout floor; the waveform is then recovered by Wiener deconvolution in a declared search band~\cite{SM}.

\begin{figure}[t]
\centering
\includegraphics[width=\columnwidth]{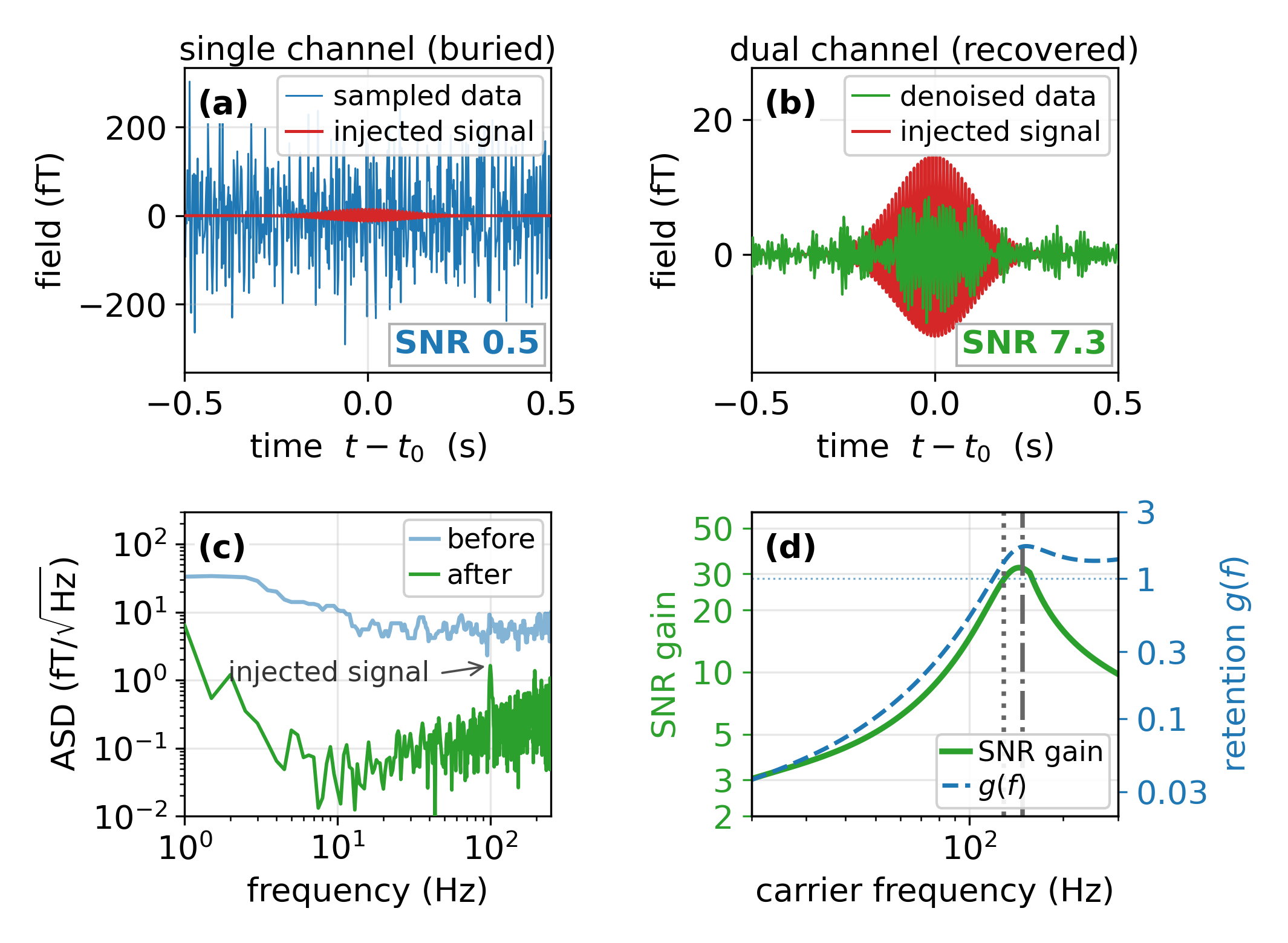}
\caption{Two-channel noise cancellation on a simulated dark-matter transient (carrier $f_0=100$\,Hz, Gaussian envelope, FWHM $0.25$\,s, $\mathcal{R}=0$); $2$\,s sampled at $500$\,Hz, in magnetic-field units. (a)~Single-channel (\K) simulation (blue) with the injected transient overlaid at true scale (red, $15$\,fT peak): buried at SNR $0.5$. (b)~\K channel after dual-beam noise cancellation, band-pass filtered to the search band as in an actual search, with the injected transient again overlaid (red); the transient emerges at SNR $7.3$, a $\times15$ improvement, its recovered amplitude reduced by the retention $g$. The SNRs are spectral quantities, fixed by the noise-floor drop and the signal retention $g\simeq0.53$, independent of this time-domain display band-pass. (c)~Amplitude spectral densities of the full records in (a) and (b): the floor near $f_0$ drops from $6.2$ (correlated noise) to $0.22\,\mathrm{fT}/\sqrt{\mathrm{Hz}}$ ($\times28$, the two-channel readout floor), while the $100$\,Hz signal peak survives (arrow). (d)~Amplitude SNR gain over the more sensitive single channel for a transient as in (a), as a function of its carrier frequency $f_0$ (green, left axis), and signal retention $g(f)$ (blue dashed, right axis); the gain peaks between the \Rb\ and \K\ antiresonances of the field-referred exotic response $\chi_j$ [Eq.~\eqref{eq:referred}] (dotted and dash-dotted verticals, $129$ and $148$\,Hz).}
\label{fig:fig3}
\end{figure}

Figure~\ref{fig:fig3} illustrates the scheme on a simulated dark-matter transient, an ALP-star crossing~\cite{JacksonKimball:2017qgk, Afach2023} in which the axion field oscillates at its Compton frequency while the transit sets the envelope, modeled here as a carrier $f_0 = 100$\,Hz under a Gaussian envelope of $0.25$\,s full width at half maximum, injected into a realistic background: shield Johnson noise at $7\,\mathrm{fT}/\sqrt{\mathrm{Hz}}$ with a $1/f$ rise below $20$\,Hz~\cite{Kornack2007,LeeRomalis2008}, an uncorrelated readout floor of $0.17\,\mathrm{fT}/\sqrt{\mathrm{Hz}}$ per channel~\cite{Kominis2003}, and light-shift and spin-projection noise within the budget of Ref.~\cite{SM}. Throughout, the noise floor denotes the rms amplitude spectral density over the declared $70$--$120$\,Hz search band (excluding $f_0\pm5$\,Hz), and the SNR is the signal's spectral amplitude at $f_0$ divided by this floor. In the single \K\ channel the transient is buried, at SNR $0.5$ [Fig.~\ref{fig:fig3}(a)]; the dual-channel subtraction cancels the correlated background, and the transient emerges in the denoised record at SNR $7.3$ [Fig.~\ref{fig:fig3}(b)], an improvement of $\times15$. The waveform itself is then recovered by Wiener deconvolution of the residual~\cite{SM}. The amplitude spectral density shows the mechanism [Fig.~\ref{fig:fig3}(c)]: the noise floor in the search band drops from $6.2$ to $0.22\,\mathrm{fT}/\sqrt{\mathrm{Hz}}$, a factor of $28$, down to the dual-channel readout floor, while the $100$\,Hz signal survives. Figure~\ref{fig:fig3}(d) generalizes the simulation across the band. Because the background suppression $S(f)$ and the signal retention $g(f)$ are frequency dependent, the net SNR gain grows with frequency across the band, reaching $\times15$ at $100$\,Hz and peaking at ${\approx}\times30$ between the \Rb\ and \K\ exotic antiresonances, where the \Rb\ exotic response $\chi_{\mathrm{Rb}}$ dips and the retention exceeds unity. At low frequencies, the referred exotic responses $\chi_j$ converge ($g \to 0$); the scheme targets the band above the detection bandwidth, where the single cell is otherwise background-limited. The suppression $S(f)$ is coupling independent and scales with the assumed correlated-to-uncorrelated noise ratio; the retention $g$ depends on $\mathcal{R}$, with the simulated $\mathcal{R} = 0$ lying within the coupling-ratio ranges predicted by the canonical axion model classes discussed below~\cite{SM}.

% -----------------------------------------------------------------------
%\section{Coupling-ratio readout}

\emph{Coupling-Ratio Readout.} Beyond detection, the dual channel measures a property of the signal that no single channel carries: the neutron-to-proton coupling ratio $\mathcal{R}$. Across the band the exotic phase difference $\Delta\varphi(\omega;\mathcal{R})$ exceeds the milliradian magnetic baseline by up to three orders of magnitude and varies steeply with $\mathcal{R}$: at $100$\,Hz it moves by hundreds of milliradians between the benchmark ratios of the canonical model classes discussed below ($-0.17$\,rad at $\mathcal{R}=0.04$, $+0.13$\,rad at $\mathcal{R}=-0.46$), so comparing the measured cross-spectral phase difference to a reference map determines $\mathcal{R}$~\cite{SM}. The map itself need not rely on the model calculation: a synthetic exotic field, formed by combining species-selective light shifts with a magnetic field~\cite{Rosenzweig2024}, generates $\Delta\varphi(\omega;\mathcal{R})$ in situ, making the readout end-to-end calibration free. The measured ratio then confronts theory. The two canonical axion model classes predict overlapping ranges: KSVZ-type models fix $\mathcal{R}=0.04(6)$, while DFSZ-type models sweep $\mathcal{R}$ monotonically across $[-0.41,+0.65]$ as $\tan\beta$ traverses its perturbative window, broadening to ${\approx}[-0.47,+0.88]$ once the coupling uncertainties are included~\cite{Kimball:2023vxk,GrillidiCortona:2015jxo,di2020landscape}. A measured $\mathcal{R}$ therefore pins down the coupling structure that any candidate model must reproduce, rather than naming the model itself: within the DFSZ class it determines $\tan\beta$; a value incompatible with $0.04(6)$ excludes the KSVZ class; a value outside ${\approx}[-0.47,+0.88]$ excludes both, pointing to a generic ALP. The axion–electron coupling present in these models can manifest in two ways. Acting directly on the electron spins, it drives the two alkalis identically and leaves \He\ untouched, so the amplitude subtraction rejects it as common mode; a field induced by the shield’s electron-spin response~\cite{jackson2016magnetic} carries the magnetic signature and is likewise rejected. In the phase channel the residual effect is negligible for KSVZ, vanishes for DFSZ-II at large $\tan\beta$ (though not at small), and for DFSZ-I modifies the phase signature at the tens-of-percent level across the band, scaled by an $\mathcal{O}(1)$ shield-response factor, so a DFSZ-I interpretation should fit the full coupling set.

A single-channel spin-based sensor is sensitive to only one effective coupling and cannot separate the proton and neutron contributions; recovering their ratio has so far required comparing several distinct devices~\cite{Rosenzweig2024}. In contrast, the calibration-free, dual-channel method can enable a measurement of the underlying coupling structure even with a single vapor cell.

% -----------------------------------------------------------------------
%\section{Discussion}

\emph{Discussion.} The correlated dual-alkali measurement adds four capabilities to a conventional comagnetometer: it subtracts the correlated magnetic background to the uncorrelated readout floor---a ${\sim}30$-fold suppression that raises the SNR by up to more than an order of magnitude (depending on coupling)---in the high-frequency band where self-compensation fails; it recognizes an exotic perturbation by an inter-species phase difference up to three orders of magnitude above the magnetic baseline; it supports a tunable antiresonance that can be brought to exact nulling; and it reads out the coupling ratio $\mathcal{R}$ calibration free from a single cell, so that for a sufficiently strong detection the same data determine the signal's nucleon coupling structure, the quantity against which candidate models are tested, information a single spin-based sensor does not carry. The antiresonance, like the phase map, can be generated and verified in situ with the synthetic exotic field~\cite{Rosenzweig2024}. The scheme applies to any signal that couples to the two species differently than a magnetic field does, and requires only modest additions to a standard search apparatus (a second probe laser, its polarimetry chain, and a dichroic splitter), allowing retrofits of existing devices.

\emph{Outlook.} The scheme is not specific to exotic fields: any perturbation that acts on the two species differently than a magnetic field does is converted into a differential signal, while the magnetic background is subtracted. This opens several directions. In principle, an electron electric dipole moment, whose atomic enhancement differs strongly between \Rb\ and \K, becomes a species-selective signal, while the magnetic noise of the electrode currents is subtracted as common mode. Dual light-shift spectroscopy at selected wavelengths can determine ratios of dipole matrix elements with high precision. The compensation optimum and the antiresonance location depend on the alkali--\He\ Fermi-contact enhancement factors, offering a single-cell precision measurement of their ratio, currently known only to a few percent~\cite{Babcock2005}. Extending the method to more alkalis with comb-referenced probes would overdetermine the couplings while suppressing differential probe drifts. This single-cell correlated approach opens a path to high-sensitivity probes of exotic fields and a new generation of detectors for precision measurements.

\emph{Acknowledgments.} We gratefully acknowledge support from the Gordon and Betty Moore Foundation, Simons Foundation, Alfred P. Sloan Foundation, and John Templeton Foundation. This work was supported in part also by the United States--Israel Binational Science Foundation Grants No.\ 2016635 and No.\ 2018257 and the Israel Innovation Authority Grants No.\ 67082 and No.\ 74482. 
YK is supported in part by the Israel Science Foundation Grant No.~1666/22. We acknowledge the use of Claude (Anthropic) for assistance with editing the manuscript, figure-preparation code, and independent numerical cross-checks of the results, and of Gemini (Google) for rendering the setup schematic of Fig.~\ref{fig:setup} from a draft drawn by the authors; all content was verified by the authors, who take full responsibility for this work.

%\bibliography{refs}  % for arXiv submission, comment this line and comment out the following line.
\onecolumngrid \vspace{1\baselineskip} \section*{Supplemental Material} 
 
\title{Supplemental Material for ``Correlated comagnetometry for precision measurements''}
 
\author{Yossi Rosenzweig}
\affiliation{Department of Physics, Ben-Gurion University of the Negev, Beer-Sheva, Israel}

\author{Yevgeny Kats}
\affiliation{Department of Physics, Ben-Gurion University of the Negev, Beer-Sheva, Israel}

\author{Eli Sarid}
\affiliation{Department of Physics, Ben-Gurion University of the Negev, Beer-Sheva, Israel}

\author{Menachem Givon}
\affiliation{Department of Physics, Ben-Gurion University of the Negev, Beer-Sheva, Israel}

\author{Yonathan Japha}
\affiliation{Department of Physics, Ben-Gurion University of the Negev, Beer-Sheva, Israel}
 
\author{Ron Folman}
\affiliation{Department of Physics, Ben-Gurion University of the Negev, Beer-Sheva, Israel}
  
\maketitle

\section{Simulation parameters and nuclear spin content}

This section collects the numerical inputs used throughout the paper. Table~\ref{tab:params} lists the cell parameters, rates, and field settings that enter the coupled Bloch equations [Eq.~(4) in the main text] for the dual-alkali \Rb-\K-\He\ system. Table~\ref{tab:spin} provides the fractional contributions of the neutron and proton spins to the total nuclear spin of each species; these spin-content factors $\sigma_{n,p}^{j}$ determine the species-dependent exotic couplings $\xi_j$ via Eq.~(5) in the main text. Table~\ref{tab:window} lists, for each alkali--noble pair, the range of the coupling ratio $\mathcal{R}\equiv\xi_n/\xi_p$ over which an exotic-field antiresonance can occur. The cell uses isotopically enriched \Rb; natural rubidium is $72\%$ $^{85}$Rb ($I=5/2$), and enrichment leaves a single rubidium isotope, so that the \Rb\ channel has a single, well-defined slowing-down factor.

\begin{table}[h!]
\caption{\label{tab:params} Simulation parameters for the dual-alkali
\Rb-\K-\He\ comagnetometer. The \He\ gyromagnetic ratio and magnetic moment
are negative; magnitudes are listed. The alkali vapor densities follow from
the saturated-vapor-pressure formulas~\cite{Seltzer2008} scaled by the
$10\!:\!90$ \Rb:\K\ mole fraction.}
\vspace{0.5\baselineskip}
\begin{tabular}{lcc}
\hline\hline
Parameter & Value & Units \\
\hline
\colrule
Temperature, $T$ & $180$ & $^{\circ}$C \\
\He\ concentration, $n_{^3\mathrm{He}}$ & $3$ & amg \\[2pt]
\Rb--\He\ spin-exchange rate coefficient, $\kappa_{\mathrm{Rb,He}}$ & $6.7\times10^{-20}$~\cite{Babcock2005} & cm$^3$/s \\
\K--\He\ spin-exchange rate coefficient, $\kappa_{\mathrm{K,He}}$ & $5.5\times10^{-20}$~\cite{Babcock2005} & cm$^3$/s \\[2pt]
\Rb--\K\ spin-exchange rate coefficient, $\kappa_{\mathrm{Rb,K}}$ & $1.2\times10^{-9}$~\cite{Babcock2005} & cm$^3$/s \\[2pt]
\Rb--\He\ enhancement factor, $\lambda_{\mathrm{Rb,He}}$ & $8\pi/3\times6.2$~\cite{Babcock2005} & \\
\K--\He\ enhancement factor, $\lambda_{\mathrm{K,He}}$ & $8\pi/3\times5.8$~\cite{Babcock2005} & \\
Electron gyromagnetic ratio, $\gamma_e$ & $2\pi\cdot2.8$ & MHz/G \\
\He\ gyromagnetic ratio, $\gamma_{^3\mathrm{He}}$ & $2\pi\cdot3.24$ & kHz/G \\
Alkali magnetic moment, $\mu_{\mathrm{alk}}$ & $9.274\times10^{-21}$ & erg/G \\
\He\ magnetic moment, $\mu_{^3\mathrm{He}}$ & $1.159\times10^{-3}\,\mu_{\mathrm{alk}}$ & erg/G \\[2pt]
\Rb\ relaxation rate, $\Gamma_{\mathrm{Rb}}$ & $600$     & Hz \\
\K\ relaxation rate, $\Gamma_{\mathrm{K}}$ & $600$ & Hz \\
\He\ relaxation rate, $\Gamma_{^3\mathrm{He}}$ & $5\times10^{-5}$ & Hz \\
\Rb\ pump rate (dual-alkali cell), $R_{p,\mathrm{Rb}}$ & $1200$ & Hz \\
\K\ pump rate (\K-\He\ cell), $R_{p,\mathrm{K}}$ & $600$ & Hz\\[2pt]
Compensation parameter, $\alpha$ & $0.578$ & \\
\Rb:\K\ ratio (mole fraction) & $10\!:\!90$ & \% \\
\Rb\ vapor density, $n_{\mathrm{Rb}}$ & $4.0\times10^{13}$ & cm$^{-3}$ \\
\K\ vapor density, $n_{\mathrm{K}}$ & $5.45\times10^{13}$ & cm$^{-3}$ \\
\Rb\ equilibrium polarization, $P_z^{\mathrm{Rb}}$ & $0.46$ & \\
\K\ equilibrium polarization, $P_z^{\mathrm{K}}$ & $0.45$ & \\
\He\ equilibrium polarization, $P_z^{\mathrm{He}}$ & $0.047$ & \\
\hline\hline
\end{tabular}
\end{table}

\begin{table}[h!]
\caption{\label{tab:spin} Fractional contributions of the neutron and
proton spins, $\sigma_n^j$ and $\sigma_p^j$, to the total nuclear spin of
each species. \He, \K\ and \Rb\ are the species of the dual-alkali cell analyzed in this work; \Ne\ is included as a further noble gas for the general comparison of Table~\ref{tab:window}. The \He\ values rest on a
full-scale shell-model calculation, the \K\ values on a detailed
perturbation-theory calculation, and the \Rb\ and \Ne\ values on the corresponding
nuclear-structure estimates; see the cited references. These spin-content
factors set the species-dependent exotic couplings $\xi_j$ through
Eq.~(5) in the main text.}
\vspace{0.5\baselineskip}
\begin{tabular}{lccc}
\hline\hline
Species & $\sigma_n$ & $\sigma_p$ & Refs. \\
\hline
\He & $0.87$ & $-0.027$ & \cite{Kimball2015,Friar:1990vx} \\
\Ne & $0.196$ & $0.013$ & \cite{Engel:1989ix} \\
\K  & $0.034$ & $-0.131$ & \cite{Kimball2015,Engel:1995gw} \\
\Rb & $0.083$ & $0.251$ & \cite{Flambaum:2006ip} \\
\hline\hline
\end{tabular}
\end{table}

\begin{table}[h!]
\caption{\label{tab:window} Range of $\mathcal{R}$ for which the read-out
alkali $A$ and the noble gas $N$ have opposite-sign exotic couplings. With
$C_j=\sigma_p^j+\mathcal{R}\,\sigma_n^j$, the condition $C_AC_N<0$ is necessary
for an exotic-field antiresonance; it holds for $\mathcal{R}$ between the
sign-reversal ratios $\mathcal{R}^*_j=-\sigma_p^j/\sigma_n^j$ when $\sigma_n^A$
and $\sigma_n^N$ share a sign (as for all pairs here), and outside that interval
when they differ. In the dual-alkali \Rb-\K-\He\ cell \K\ has both partners, so
its range is the union $(-3.02,3.85)$, which contains $\mathcal{R}=0$.}
\vspace{0.5\baselineskip}
\begin{tabular}{lcc}
\hline\hline
Pair ($A$--$N$) & $(\mathcal{R}^*_A,\,\mathcal{R}^*_N)$ & Antiresonance range \\
\hline
\K--\He  & $(3.85,\ 0.03)$   & $0.03<\mathcal{R}<3.85$ \\
\Rb--\He & $(-3.02,\ 0.03)$  & $-3.02<\mathcal{R}<0.03$ \\
\K--\Ne  & $(3.85,\ -0.07)$  & $-0.07<\mathcal{R}<3.85$ \\
\Rb--\Ne & $(-3.02,\ -0.07)$ & $-3.02<\mathcal{R}<-0.07$ \\
\hline\hline
\end{tabular}
\end{table}

\section{The compensation optimum $\alpha^\star$}

The compensation field of Eq.~(6) in the main text interpolates only the \He-on-alkali term between the Rb-optimal ($\alpha=0$) and K-optimal ($\alpha=1$) settings, because a single applied $B_z$ cannot simultaneously compensate both alkalis when $\lambda_{\mathrm{Rb,He}} \neq \lambda_{\mathrm{K,He}}$. The optimal weight has a simple physical origin. The \Rb--\K\ spin-exchange rate is large (Table~\ref{tab:params}), so the two alkalis are locked into a single collective transverse mode and precess together. This collective spin experiences the \He\ magnetization not as two separate fields but as one effective field equal to the number-weighted average of the individual couplings,
\begin{equation}
\langle \lambda_{A,\mathrm{He}} \rangle
= \frac{n_{\mathrm{Rb}}\,\lambda_{\mathrm{Rb,He}} + n_{\mathrm{K}}\,\lambda_{\mathrm{K,He}}}{n_{\mathrm{Rb}} + n_{\mathrm{K}}} ,
\end{equation}
where the weights are the polarized-spin populations of the two alkalis (equal here, since $P_z^{\mathrm{Rb}} \simeq P_z^{\mathrm{K}}$, so the number densities suffice). Matching the interpolated \He\ term $[(1-\alpha)\lambda_{\mathrm{Rb,He}} + \alpha\lambda_{\mathrm{K,He}}]$ to this collective coupling gives
\begin{equation}
\alpha^\star = \frac{n_{\mathrm{K}}}{n_{\mathrm{Rb}} + n_{\mathrm{K}}} \approx 0.576 ,
\label{eq:alpha_star}
\end{equation}
the \K\ fraction of the total alkali density. For the cell of Table~\ref{tab:params}, Eq.~\eqref{eq:alpha_star} evaluates to $0.576$, and the value that numerically minimizes the residual very-low-frequency (DC) magnetic response is $0.579$ for the \K\ channel ($0.570$ for \Rb), bracketing Eq.~\eqref{eq:alpha_star}. At this weight a DC magnetic field is suppressed by ${\sim}1.4\times10^{4}$ in the \K\ channel and ${\sim}2.7\times10^{3}$ in the \Rb\ channel relative to the on-resonance response---one to two orders of magnitude better than the ${\sim}150$ obtained when either alkali is set to its own compensation point ($\alpha=0$ or $1$), because the compromise matches the locked collective mode rather than either alkali individually. We adopt $\alpha = 0.578$ throughout.

Compensation nulls the DC \emph{amplitude} response, but the two alkalis keep a small difference in Larmor frequency, which leaves a residual magnetic phase difference $\Delta\varphi^B$ that grows toward low frequency: about $2.5$\,mrad at $100$\,Hz, $15$\,mrad at $10$\,Hz, and larger toward DC. The phase-based rejection therefore stays at the milliradian level within the $70$--$120$\,Hz search band considered here, and degrades below it; equalizing the two Larmor frequencies with a species-selective light shift would extend it to lower frequencies.

\section{The exact-null locus of the antiresonance}

The transverse polarization is decomposed in the Bloch ansatz $P_\perp(t) = P_+ e^{i\omega t} + P_- e^{-i\omega t}$, with $P_+$ and $P_-$ obtained from the linearized steady state. The $x$-projection is $P_x = [P_+ + (P_-)^*]/2$, which is proportional to the observable [Eq.~(2) of the main text], and vanishes exactly when $P_+ = -(P_-)^*$. This is a single complex condition, equivalent to two real ones: amplitude equality, $|P_+| = |(P_-)^*|$, and a phase offset of $\pi$ between $P_+$ and $(P_-)^*$. These constitute two real equations on the three-parameter space $(\alpha, \mathcal{R}, \omega)$, so the subset on which both hold simultaneously is one-dimensional, and along it $|P_x^\mathrm{K}|$ is exactly zero. The operating points shown in Fig.~2(b) of the main text illustrate the sharp difference between antiresonances that do ($\alpha=0.773$, $72$\,Hz) and do not ($\alpha=0.578$, $157$\,Hz) belong to this subset. The same mechanism produces an analogous antiresonance in the \Rb\ response at a different frequency ($\approx130$\,Hz at $\alpha=0.578$, $\mathcal{R}=0$), which is exploited by the amplitude channel in Fig.~3(d) of the main text.

\section{The phase-difference map and recovery of the coupling ratio}

\begin{figure}[h!]
\centering
\includegraphics[width=0.9\linewidth]{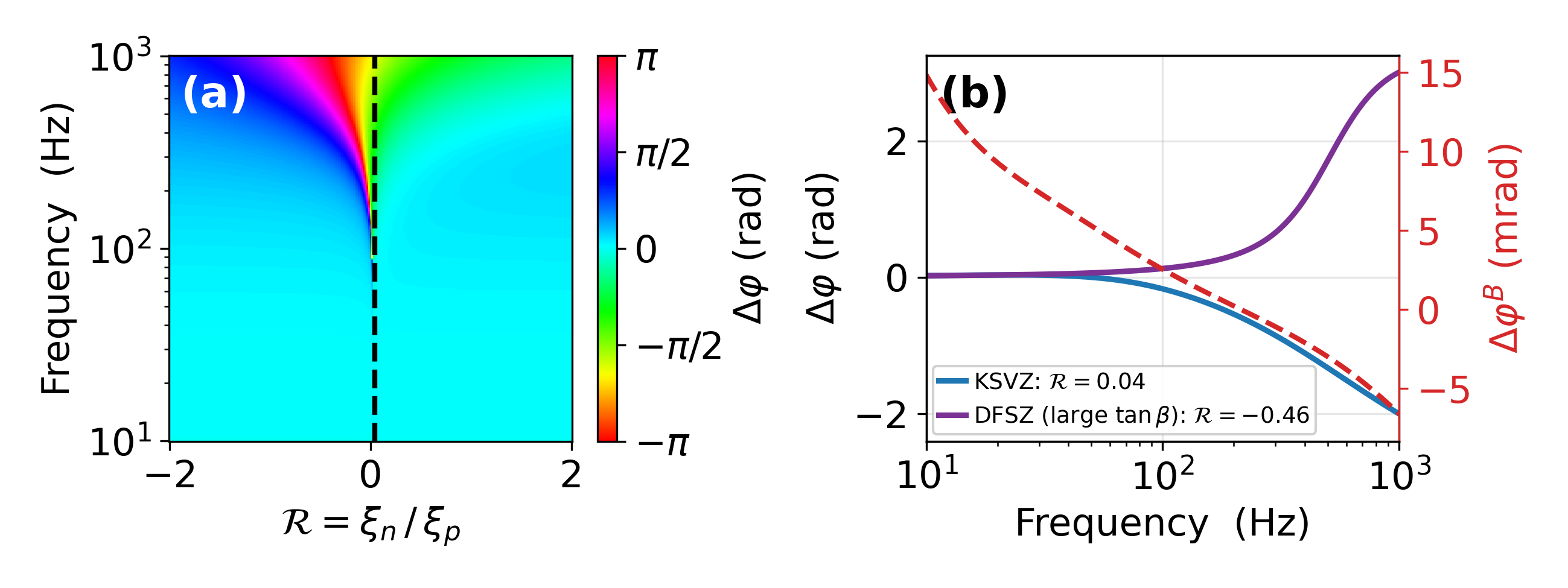}
\caption{\label{fig:smdphi}$\Delta\varphi(\omega; \mathcal{R})$ for the dual-alkali Rb-K-\He\ cell at $\alpha = 0.578$. (a) Heatmap of $\Delta\varphi$ across the $(\mathcal{R}, \omega)$ plane, wrapped to $[-\pi, \pi)$ for the cyclic color bar; the black dashed line marks the central value of the KSVZ prediction, $\mathcal{R} = 0.04$. (b) Unwrapped cross sections at two representative model-motivated ratios, $\mathcal{R} = 0.04$ (central KSVZ prediction, blue) and $\mathcal{R} = -0.46$ (large-$\tan\beta$ DFSZ boundary, purple), comparing the exotic phase difference $\Delta\varphi$ (left axis, radians) to the magnetic baseline $\Delta\varphi^B$ (dashed red, right axis, milliradians): a separation of up to three orders of magnitude enabling high common-mode rejection. Beyond the antiresonance $|\Delta\varphi|$ exceeds $\pi$, so the unwrapped curves in (b) equal the wrapped map in (a) up to $2\pi$. The map suppresses the magnetic response to the milliradian level while serving as a matched filter for a specific coupling ratio $\mathcal{R}$.}
\end{figure}

Figure~\ref{fig:smdphi} shows how to recover $\mathcal{R}$ by mapping $\Delta\varphi$ across the $(\omega, \mathcal{R})$ plane [panel (a)] and contrasts the exotic and magnetic responses along representative cross sections [panel (b)]: for the model benchmarks the exotic phase difference exceeds the magnetic baseline $\Delta\varphi^B$ by up to three orders of magnitude across the band, while $\Delta\varphi^B$ itself remains negligible across the band. Given a candidate transient that triggers the system, one computes the cross-spectral phase between the simultaneously recorded $P_x^{\mathrm{K}}(t)$ and $P_x^{\mathrm{Rb}}(t)$ records over the frequency band and compares it to the calibration map $\Delta\varphi(\omega;\mathcal{R})$ for $\mathcal{R}$. 

The nuclear spin-content factors $\sigma^j_{n,p}$ carry theory uncertainties: at the percent level in absolute terms for \He\ and larger for the alkalis. For the recovered $\mathcal{R}$ this translates to a comparable fractional uncertainty. The sign-reversal ratio $\mathcal{R}^\star_{\mathrm{He}} = -\sigma_p^{\mathrm{He}}/\sigma_n^{\mathrm{He}} = 0.027/0.87 \approx +0.03$ is more sensitive, since $\sigma_p^{\mathrm{He}}$ is itself a small number: a percent-level absolute shift moves $\mathcal{R}^\star_{\mathrm{He}}$ by tens of percent, shifting the nucleon ratio to which the measured antiresonance and benchmark phases correspond. These spin-content uncertainties enter the recovered $\mathcal{R}$ on the same footing as the theory uncertainties of the model predictions themselves: KSVZ-type models fix $\mathcal{R} = 0.04 \pm 0.06$, dominated by the nucleon coupling $C_n = -0.02(3)$, with $C_p = -0.47(3)$~\cite{GrillidiCortona:2015jxo}, while DFSZ-type models sweep $\mathcal{R}$ monotonically downward across $[-0.41, +0.65]$ as $\tan\beta$ (defined as $v_u/v_d$) traverses its perturbative window $0.25 \le \tan\beta \le 170$, broadening to ${\approx}[-0.47, +0.88]$ once the coupling uncertainties are included, and passing through the KSVZ band near $\tan\beta \approx 0.74$. The observable is therefore reported as a measured coupling ratio: it excludes a model class only when the value falls outside that class's range, and it cannot separate the classes where their predictions coincide. 

Cross-talk between the two probe channels contributes a small systematic to $\Delta\varphi$. Optical leakage of one probe into the other's photodiode is suppressed to the $10^{-6}$ level by a narrow-band interference filter in each polarimetry arm. What remains is atomic: each probe picks up a small Faraday rotation from the other alkali's far dispersive tail, at the percent level or below for probes detuned ${\sim}0.1$\,THz from their own D1 lines, which translates to a $\Delta\varphi$ offset of a few mrad. This offset is static and is absorbed by the in-situ synthetic-field calibration. A strong light-shift transient might appear as an exotic-field candidate: it does not share the magnetic channel ratio, so it survives the correlated subtraction and shows up in the residual $r$ just as a genuine signal would. The second observable, the phase difference, separates the two. A light shift couples to the alkalis species-selectively rather than through the common electronic gyromagnetic ratio, so it arrives with its own phase signature, set by the ratio of the shifts it imposes on the two alkalis. This ratio is fixed by the laser detunings and does not depend on the amplitude of the transient. In the Bloch model at $100$\,Hz, a shift acting on Rb (K) alone produces $\Delta\varphi = -66$ ($+51$)\,mrad, compared with $+2.5$\,mrad for the magnetic background and $-0.17$ ($+0.13$)\,rad for an exotic drive at $\mathcal{R} = 0.04$ ($-0.46$). A candidate whose phase difference falls off the calibrated map $\Delta\varphi(\omega;\mathcal{R})$ is therefore identified as a light-shift artifact and rejected, at no cost to genuine events.

\section{Two-channel cancellation and Wiener recovery}

The field-referred channels of Eq.~(7) in the main text, $x_j = \tilde{\theta}_F^j/T_j^B$, are combined as $r(\omega) = x_{\mathrm{K}}(\omega) - \hat{H}(\omega)\, x_{\mathrm{Rb}}(\omega)$, with the transfer estimate
\begin{equation}
\hat{H}(\omega) = \frac{\langle x_{\mathrm{K}}(\omega)\, x_{\mathrm{Rb}}^*(\omega)\rangle}{\langle |x_{\mathrm{Rb}}(\omega)|^2\rangle}
\label{eq:H}
\end{equation}
computed by Welch averaging over exotic-signal-free records (the magnetic background is always present and is precisely what $\hat{H}$ learns); for transient searches these are the off-trigger records, while for a persistent (continuous) signal $\hat{H}$ must be estimated from neighboring frequency bins to avoid partial self-nulling of the signal. The quality of $\hat{H}$ is set by the number of averaged segments rather than by any single window: in the band where the correlated background dominates the referred channels, $x_j$, the ordinary coherence between them is near unity, so the cross-spectral estimator converges within a handful of segments (seconds of data at the simulation sampling), and the useful record length is limited by the stationarity of the background rather than by counting statistics. The simulation estimates $\hat{H}$ from a $200$\,s signal-free record ($100$ two-second segments). Since the correlated background dominates both referred channels, $\hat{H} \to 1$, and any miscalibration of the magnetic transfer functions $T^B_j$ folds into $\hat{H}$ and cancels; the model enters only the amplitude scale of the recovered waveform. The exotic waveform is then recovered by Wiener deconvolution in the declared search band,
\begin{equation}
\hat{b}(\omega) = \frac{G^*(\omega)\, r(\omega)}{|G(\omega)|^2 + S_{nn}(\omega)/S_b(\omega)}, \qquad G \equiv \chi_{\mathrm{K}} - \hat{H}\,\chi_{\mathrm{Rb}},
\label{eq:wiener}
\end{equation}
where $S_{nn}$ is the residual (uncorrelated) noise spectrum and $S_b$ a flat prior for the signal spectrum in the band; no template for the signal's shape or timing is assumed. The simulation of Fig.~3 in the main text uses a sampling rate of $500$\,Hz, a declared search band of $70$--$120$\,Hz, and a $2$\,s spectral window, with the transient amplitude set so that the single-channel spectral SNR is 0.5.

The search itself is broadband: the subtraction and the retention $g(f)$ operate across the declared band at a fixed operating point, with no scanning, and the usual per-bin trials factor of a spectral search applies. The phase readout requires no tuning: the cross-spectral phase of the recorded data is compared to the map at the known operating point $\alpha$. It is possible to choose the operating point in advance to place the steep region of the map within the search band; for persistent candidates, stepping $B_z$ additionally tests whether the signal tracks the calculated $\alpha$ dependence.

\section{Noise budget for the dual-channel subtraction}

The adaptive subtraction removes any perturbation that carries a common magnetic signature; the residual floor is therefore set by noise that is not common-mode. Four classes are relevant.

\emph{Species-selective (light-shift) noise.} Pump light-shift fluctuations, driven by drifts in probe intensity, detuning, or beam pointing, act on a single alkali as an effective transverse field, the same signature class as the exotic signal. The spin-exchange locking that pins the magnetic phase difference also transmits a \Rb-only effective field to the \K\ channel with amplitude ratio $0.975$ at $100$\,Hz; accounting for the small relative phase of the two field-referred transfers, the subtraction suppresses it by ${\sim}14$, leaving ${\approx}7\%$. The exotic signal is not suppressed in the same way because part of it is carried by a \He-mediated arm that a light shift lacks. An effective-field drive appears in the field-referred channel with transfer $|T^{\mathrm{LS}}_{\mathrm{Rb}}/T^{B}_{\mathrm{Rb}}| \approx 0.43$ at $100$\,Hz; requiring the residual to fall below the recovered $0.22\,\mathrm{fT}/\sqrt{\mathrm{Hz}}$ floor then sets a budget line $\delta B_{\mathrm{LS}} \lesssim 7\,\mathrm{fT}/\sqrt{\mathrm{Hz}}$ of \Rb-equivalent effective field at $100$\,Hz, met by standard detuning and intensity control. The simulation of Fig.~3 injects $1\,\mathrm{fT}/\sqrt{\mathrm{Hz}}$, well within this budget.

\emph{Response-ratio stability.} The cancellation holds while the ratio $c \equiv T^B_{\mathrm{K}}/T^B_{\mathrm{Rb}}$ of the two magnetic transfer functions is stable. A fractional drift $\delta c$ in this ratio leaks the correlated background into the residual $r$ at the level $\delta c$, so reaching the $0.22\,\mathrm{fT}/\sqrt{\mathrm{Hz}}$ floor requires $\delta c \lesssim 3\%$ over the timescale on which $\hat{H}$ is re-estimated: This is a stability requirement, not a calibration: $\hat{H}$ measures the ratio from the data, so its absolute value is never required; it must only stay constant between successive re-estimations. This requirement is loose: because spin-exchange locking makes the two alkalis respond as a single collective mode, $c$ is nearly insensitive to the operating point---in the Bloch model, $d\ln|c|/dT \approx 5\times10^{-4}\,\mathrm{K}^{-1}$ at $100$\,Hz, so ordinary cell-temperature control leaves orders of magnitude of margin, and $\hat{H}$ is in any case re-estimated from each ${\sim}200$\,s record. The operating point itself, and with it the antiresonance location and the phase map, follows the \He\ magnetization through the compensation condition. Closed-loop control of the compensation field, demonstrated on the same \K--\Rb--\He\ platform in Ref.~\cite{Klinger2023}, holds this point fixed against the slow (hours-scale) \He\ polarization dynamics, so the phase map needs only occasional re-verification with the synthetic-field calibration rather than continuous re-measurement.

\emph{Differential gradient sampling.} \Rb\ is pumped directly while \K\ is polarized through spin exchange, so the two polarization profiles are not identical, the two channels sample the field at slightly different effective positions, and a magnetic-field gradient is not perfectly common-mode. The offset between the two effective positions is bounded by the distance a K atom diffuses before equilibrating with the local Rb polarization, $\ell = \sqrt{q D / (k_{\mathrm{RbK}}\, n_{\mathrm{Rb}})} \approx 50\,\mu\mathrm{m}$ for the parameters of Table~\ref{tab:params}, so the two profiles coincide to this scale. Even allowing a tenfold margin ($0.5$\,mm), shield Johnson-noise gradients at the level of Ref.~\cite{Kornack2007} contribute at most $0.08\,\mathrm{fT}/\sqrt{\mathrm{Hz}}$, a fraction of the residual floor.

\emph{Spin-projection noise.} The standard single-species estimate, $\delta B \approx (1/\gamma_e)\sqrt{2 q \Gamma / (n_j V)}$~\cite{Allred2002,Kominis2003}, with the densities of Table~\ref{tab:params} and the effective volume $V \approx 1\,\mathrm{cm}^3$ over which the $10$\,mm probe beam overlaps the pumped column in our $20$-mm-diameter spherical cell, gives $0.06$ and $0.07\,\mathrm{fT}/\sqrt{\mathrm{Hz}}$ for the \K\ and \Rb\ channels; added in quadrature by the subtraction, this is about $0.09\,\mathrm{fT}/\sqrt{\mathrm{Hz}}$, a factor of two below the readout floor.

All four error budget terms lie at or below the recovered $0.22\,\mathrm{fT}/\sqrt{\mathrm{Hz}}$ readout floor, so none limits the simulation of Fig.~3 in the main text.

\bibliography{refs}

\end{document}